\documentclass[12pt,a4paper]{JHEP3}
\usepackage{amsmath,amssymb,bm}
\usepackage{latexsym}
\usepackage{graphicx}
\usepackage{cite}

\def\eqref#1{Eq.~(\ref{#1})}

\def\Eq#1{\begin{equation} #1 \end{equation}}

\newcommand{\pd}{\partial}

\def\X5sp{{\rm X}_5}
\def\Y3sp{{\rm Y}_3}
\def\Z3sp{{\rm Z}_3}

\newcommand{\bea}{\begin{eqnarray}}
\newcommand{\eea}{\end{eqnarray}}
\newcommand{\beq}{\begin{equation}}
\newcommand{\eeq}{\end{equation}}

\title{
Compact objects and the swampland
}
\author{
Carlos A. R. Herdeiro\\
CENTRA, Departamento de F\'isica, Instituto Superior T\'ecnico,\\
~~Universidade de Lisboa, Avenida Rovisco Pais 1, 1049 Lisboa, Portugal.}
\author{
Eugen Radu\\
School of Theoretical Physics, Dublin Institute for Advanced Studies, 
 10 Burlington Road, Dublin 4, Ireland and
\\
Center for Research and Development in Mathematics and Applications, \\
Campus de Santiago, 3810-183 Aveiro, Portugal.}
\author{
Kunihito Uzawa\\
Department of Physics,
School of Science and Technology,\\
~~Kwansei Gakuin University, Sanda, Hyogo 669-1337, Japan.}
\abstract{%
Recently, two simple criteria were proposed to assess if vacua emerging from an effective scalar field theory are part of the string ``landscape" or ``swampland". The former are the vacua  that emerge from string compactifications; the latter are not obtained by any such compactification and hence may not survive in a UV completed theory of gravity.  So far, these criteria have been applied to inflationary and dark energy models. Here we consider them in the context of solitonic compact objects made up of scalar fields: boson stars. Analysing several models (static, rotating, with and without self-interactions), we find that, in this context, the criteria are not independent. Furthermore, we find the universal behaviour that in the region wherein the boson stars are expected to be perturbatively stable, the compact objects may be part of the landscape. By contrast, in the region where they may be faithful black hole mimickers, in the sense they possess a light ring, the criteria fail (are obeyed) for static (rotating) \textit{ultracompact} boson stars, which should thus be part of the swampland (landscape). We also consider hairy black holes interpolating between these boson stars and the Kerr solution and establish the part of the domain of existence where the swampland criteria are violated.  In interpreting these results one should bear in mind, however, that the swampland criteria are not quantitatively strict.
}

\begin{document}

\section{Introduction and motivation}
 \label{sec:motivation}
Theoretical consistency and observational data require two epochs of accelerated expansion to be present in a successful cosmological model: early universe inflation and present day acceleration 
\cite{Riess:1998cb}.
It then becomes a pressing task to embed such cosmological model in fundamental physics, in particular within a consistent theory of quantum gravity. Such research program has been under intense scrutiny in string theory and its low energy limit supergravity (see $e.g.$ \cite{Grana:2005jc, Douglas:2006es, Blumenhagen:2006ci} for reviews), the most developed approach to quantum gravity.  
A main obstacle is the fact that these theories require spacetime to be higher dimensional. Then, in order to make contact with low energy four dimensional physics,  a natural way to conceal the extra dimensions must be found, usually via a \textit{compactification}. This compactification, however, gives rise to new problems. 
A most serious one is the issue of moduli stabilization 
\cite{Gukov:1999ya,Kachru:2003aw, Kachru:2003sx, Balasubramanian:2005zx}. 
Another issue is the no-go theorems against accelerating and expanding 
universes in simple Kaluza-Klein-type or stationary warped
compactifications, with a smooth compact internal space 
\cite{deWit:1986mwo,Maldacena:2000mw,Ivanov:2000fg}. 
  Yet, despite considerable efforts, de Sitter vacua found in string theory are, at best, meta-stable 
\cite{Goheer:2002vf, Kachru:2003aw, Kachru:2003sx,Danielsson:2018ztv}. 
 An emerging view, at present, is that %
fully-stable de Sitter vacua may actually not exist in String Theory 
\cite{Danielsson:2018ztv}. 

In string theory model building, on the other hand, a huge number of possibilities exists, coming from the choice of various ingredients, including the 
compactification manifold, background fluxes and the presence of different types of branes. This remarkably 
large space of inequivalent string backgrounds is called the 
\textit{string landscape} 
\cite{Susskind:2003kw,Vafa:2005ui, Denef:2008wq}. 
 At present, it is unknown how to identify which, if any,  
particular choice among the enormous set of possibilities, 
describes our universe.  This large number of string vacua, together with the inability to find appropriate selection mechanisms, has made \textit{top-down} approaches to identify the correct vacuum  unsuccessful.  Consequently, the research programme towards identifying the correct string vacuum has shifted towards a \textit{bottom-up} approach over the past decade \cite{Brennan:2017rbf}. That is, instead of starting with a ten-dimensional string theory and study its compactifications down to four-dimensional effective theories, one starts studying four-dimensional quantum field theories and trying to couple them to gravity.  Such ``bottom-up" approach, however, faces a key question: does any consistent looking four-dimensional effective field theory coupled to gravity  arise, in some way, from a string compactification? If so, since string theory is a candidate for an ultra-violet (UV) complete theory of gravity, such  consistent looking effective field theory may indeed admit a UV completion. Otherwise, the theory may not have a UV completion as a quantum theory of gravity \cite{Brennan:2017rbf}. To distinguish the latter from the string landscape, such class of theories is said to form the \textit{swampland}~\cite{Vafa:2005ui, Ooguri:2006in}.  

It thus becomes crucial to identify a set of ``swampland criteria" which identify if an effective field theory admits a string theory UV completion or not. A precise derivation of such criteria is, obviously, challenging. Nonetheless,  a proposal of simple and minimal criteria for  an effective \textit{scalar field theory} to be \textit{safe} from the swampland was put forward in
\cite{Ooguri:2016pdq, Obied:2018sgi, Agrawal:2018own, Ooguri:2018wrx}.
 These criteria enforce the absence of quantum correction and the hypothesis that de Sitter (or sufficiently close to it) vacua do not exist in the string landscape, 
and have  been applied to various inflationary models recently, see $e.g$~\cite{Brown:2015iha}. 
  Although swampland conjectures which exclude de Sitter vacua in string theory have been much discussed, little attention has been given to examine the swampland criteria in the context of solitonic compact objects made up of scalar fields.
	Our motivation for the present work is to improve this situation. 

Some scalar field theories allow the existence of solitonic, self-gravitating, compact objects 
dubbed \textit{boson stars}~\cite{Schunck:2003kk}. 
These localised lumps of energy have been widely studied since their discovery half a century ago~\cite{Kaup:1968zz,Ruffini:1969qy}. Some boson stars have good dynamical properties: they can form dynamically~\cite{Seidel:1993zk} 
and are perturbatively stable
~\cite{Jetzer:1988vr}. 
They have also interesting phenomenological properties: boson stars have been proposed as black hole mimickers 
(see $e.g.$~\cite{Guzman:2009zz,Cunha:2017wao}) and dark matter candidates
\cite{Suarez:2013iw}. Their appearance in the context of string compactifications has been studied in~\cite{Krippendorf:2018tei}. 
Moreover, rotating boson stars~\cite{Schunck:1996he} 
belong to a larger family of solutions of a (massive, complex) scalar field theory coupled to gravity that contains hairy black holes - 
an equilibrium bound state of a horizon with a rotating boson star, 
in synchronous rotation~\cite{Herdeiro:2014goa,Herdeiro:2015gia}. Such family of hairy black holes interpolates between the solitonic boson stars, and (in the vanishing scalar field limit) the paradigmatic Kerr black hole of general relativity~\cite{Kerr:1963ud}. 

In this work we shall apply the recently proposed swampland criteria to these compact objects made up of scalar fields, which are vacua in some scalar field theories. Loosely speaking these objects may be dilute, compact or ultra-compact. The latter can bend light into bound orbits, and in particular possess~\textit{light rings}. One would expect that very compact objects may be inconsistent with the swampland criteria. We test this intuition by considering both spherical boson stars and rotating ones, with~\cite{Colpi:1986ye} and without self-interactions and both fundamental and excited boson stars, as well as hairy black holes. Our main conclusion is universal: the swampland criteria only fail in the branch of solutions which are expected to be unstable, which roughly corresponds to where they are the most compact. Another observation is that when the solutions are ultra-compact, and thus have the potential to be faithful 
black hole mimickers~\cite{Cardoso:2016rao}, 
the swampland criteria may or may not fail: they fail for the spherical models we have analysed (all spherical boson stars) but they are obeyed by some ultra-compact \textit{rotating} boson stars.  
We also analyse the swampland criteria for hairy black holes interpolating between rotating boson stars and the Kerr solution and exhibit the region in parameter space where the criteria are violated, which roughly correspond to the region where the scalar hair is the most compact around the horizon. Another novel observation in this application is that the two criteria are not independent. As a word of caution concerning all these results, the swampland criteria are not defined as providing strict inequalities, but rather estimates. This should be considered in interpreting the results herein. 

This paper is organised as follows. In Section~\ref{sec:criteria} we discuss the swampland criteria. In Section~\ref{sec:compact} we present the compact objects we wish to consider and apply the swampland criteria. In Section~\ref{sec:discussion} we discuss our results and their implications.

%
\section{Swampland criteria} 
\label{sec:criteria}

String theory provides a huge landscape of vacua. Such landscape is populated by consistent low energy effective theories and is surrounded by a region where inconsistent semiclassical effective theories live: the swampland~\cite{Vafa:2005ui, Ooguri:2006in}.  A significant portion of effective field theories may fall into the swampland. 

Two criteria have been proposed to test if a given configuration \textit{is safe} from quantum corrections within a consistent theory of quantum gravity 
 \cite{Dasgupta:2018rtp}. 
These are called the \textit{swampland criteria}. 
Following~\cite{Agrawal:2018own} and adapting to our case of interest these criteria are the following.

Consider a theory of quantum gravity coupled to two scalar fields, $\phi^i$, $i=1,2$. The effective Lagrangian, valid up to some cutoff scale, reads:

\begin{eqnarray}
\label{L}
\mathcal{L}=\frac{R}{2}-\frac{1}{2}g^{\mu\nu}\partial_\mu \phi^1\partial_\nu \phi^1-\frac{1}{2}g^{\mu\nu}\partial_\mu \phi^2\partial_\nu \phi^2 -V(\phi^1,\phi^2) + \dots \  ,
\end{eqnarray}
where Planck units, $M_p=\sqrt{8\pi G}=1$, are used.

\begin{description}
\item[Swampland criterion 1:] 
The range of the scalar field has an upper bound 
  \cite{Ooguri:2006in, Grimm:2018ohb}:
\begin{eqnarray}
\Delta \phi^i \sim \mathcal{O}(1) ,
\end{eqnarray}
where $\Delta$ stands for the difference between the maximal and minimal value of the scalar field. 
\item[Swampland criterion 2:] There is a lower bound on the relative variation of the potential in field space:
\begin{eqnarray}
\label{crit2}
\frac{|\nabla_\phi V|}{V}>c\sim \mathcal{O}(1) \ ,
\end{eqnarray}
\end{description}
where $c$ is a positive constant of order 1 and one should take 
$\left|\nabla V\right|=\sqrt{\delta^{ij}\pd_{\phi^i}V\pd_{\phi^j}V}$\,.

Criterion 1 signifies that the scalar field in the low energy 
effective field theory is only valid over a field displacement~\cite{Ooguri:2016pdq} in Planck units, 
measured as a distance in the target space geometry. Then, scalar field excursions in field space are bounded. The rationale of this criterion is that when a large distance $D\gg 1$ in field space is explored, 
a tower of light modes appears with mass scale
\Eq{
m\sim M_p\exp\left(-\alpha D\right)\,,
}
where the parameter $\alpha$ is given by 
$\alpha\sim {\cal O}(1)$\,, and $M_p$ denotes the 
four-dimensional Planck mass. 
These modes invalidate the above four-dimensional 
effective action.

The second swampland criterion is motivated by the observation 
that it appears to be difficult to construct any reliable 
de Sitter vacua in string theory -- see arguments and no-go theorems in \cite{Brennan:2017rbf, Obied:2018sgi}. Then, any theory with the effective action 
(\ref{L}) that is a low energy effective theory 
for a consistent quantum gravity theory should verify the 
criterion (\ref{crit2}). This has the crucial implication 
that the four-dimensional theory should not admit a de Sitter solution; in other words,  the value of the 
scalar potential at an extremum must be non-positive. Then it only allows 
for Minkowski or anti-de Sitter solutions amongst maximally symmetric spacetimes.  If true, this swampland criteria has profound observational implications, namely that dark energy must vary sufficiently fast in time -- it is not a cosmological constant. 

Let us remark that recently, a refined version of de Sitter conjecture has been proposed \cite{Ooguri:2018wrx} (see~\cite{Andriot:2018mav} for yet another refinement). It states that either (\ref{crit2}) holds
or the minimum of the Hessian is bounded 
\Eq{
{\rm mim}\left(\nabla_i\nabla_j V\right)\le -c'\,V\,, 
}
where $c'$ is positive and of order one in Planck units. The refined criterion turns out to evade several 
counter-examples 
\cite{Denef:2018etk}, but a potential counter-example to the refined version is suggested in  
\cite{Blaback:2018hdo}. Here we shall use the two criteria presented above.

%
\section{Application to boson stars and hairy black holes} 
\label{sec:compact}

To apply the above swampland criteria to a class of compact objects we shall restrict ourselves to the case of two real fields, \textit{cf.}~\eqref{L}, with the same mass and a specific self-interaction. The two real fields in (\ref{L}) combine to give a massive, complex scalar field
\begin{eqnarray} 
\label{psi}
\psi\equiv \phi^1+i \phi^2  \ ,
\end{eqnarray}
and the Lagrangian (\ref{L})  becomes equivalent to
\begin{eqnarray}
\label{newL}
\mathcal{L}=\frac{R}{2}- \frac{1}{2}g^{\alpha\beta} \partial_\alpha \psi \partial_\beta \psi^*  - V( |\psi| ) \ , 
\qquad V( |\psi| )=\frac{\mu^2}{2} |\psi|^2 + \frac{\lambda}{4} |\psi|^4 \ ,
\end{eqnarray}
with $|\psi|^2=\psi \psi^*$. We have chosen the potential to include a mass term (mass $\mu$) plus a positive quartic self-interacting term (coupling $\lambda$). For $\lambda=0$, the boson star solutions have a maximal mass that scales as $\sim M_p^2/\mu$. This implies that for boson stars to achieve solar masses the scalar field mass must be ultralight, $\mu \sim 10^{-10}$ eV. Although such ultralight particles are predicted to exist in beyond the Standard Model physics, e.g., the string axiverse~\cite{Arvanitaki:2009fg}, they are not present in the Standard Model. The inclusion of the quartic self-interaction changes the scaling of the maximal mass to $\sim M_p^3/\mu^2$ (see~\cite{Herdeiro:2015tia} for an extended discussion and a list of references), making boson stars with astrophysical (solar) masses compatible with standard model order masses for the elementary bosonic particles. This is the main motivation to consider the potential (\ref{newL}), 
which, from a quantum field theory viewpoint, is also renormalizable. 

Then the two criteria, in terms of the complex scalar field $\psi$, become:
\begin{eqnarray}
{\bf Criterion \ 1 }: \Delta |\psi| \lesssim  1 \ , \qquad  {\bf Criterion \ 2 }: \ 
\left| \frac{ \partial V}{\partial |\psi|} \right | \frac{1}{V(|\psi|)}
\gtrsim 1 \ .
 \end{eqnarray}
The second criterion, becomes, explicitly, in terms of the chosen potential,
\begin{eqnarray}
  \left|\frac{ \partial V}{\partial |\psi|} \right | \frac{1}{V(|\psi|)}
	=\frac{2}{|\psi|}
	\frac{1+\frac{\lambda |\psi|^2}{\mu^2}}{1+\frac{ \lambda |\psi|^2}{2\mu^2}}
	\gtrsim 1 \ .
 \end{eqnarray}
Thus,  for the simplest case, $\lambda=0$ (no self-interaction),  ${\bf Criterion \ 2 }$ translates into 
\begin{eqnarray}
 |\psi|  
\lesssim 2 \ .
\label{c1}
 \end{eqnarray}
 It follows that all solutions which satisfy ${\bf Criterion \ 1 }$,
$i.e.$
\begin{eqnarray}
\Delta |\psi| \lesssim  1
\label{c2}
 \end{eqnarray}
will satisfy also ${\bf Criterion \ 2 }$ (as $\Psi$ vanishes at spatial infinity for asymptotically flat boson stars or hairy black holes). In these relations, we interpret 
$\Delta |\psi|$
as the maximal variation of the scalar field in a given configuration.

In the general case with self-interactions, $\lambda> 0$, observing that
\begin{eqnarray}
2\geqslant \frac{1+\frac{\lambda |\psi|^2}{\mu^2}}{1+\frac{ \lambda |\psi|^2}{2\mu^2}} \geqslant 1 \ ,
 \end{eqnarray}
one concludes that~\eqref{c2} implies Criterion 2. Thus our analysis reduces to analysing condition~\eqref{c2}. It is a novel feature of the vacua we are analysing that the two criteria are not independent.
 
The dependence between the two criteria we have just described may look intriguing. For boson stars there can be non-negligible energy due to the spatial  gradient of the fields. Then, the kinetic term of the scalar field might be comparable to the scalar potential  at some points. Hence, it may be unclear why criterion 2 is implied by criterion 1. The reason is that even though boson stars have 
a non-negligible kinetic energy, the gradient terms always relate to the 
overall variation of the scalar field. This is clearest for spherical 
boson stars, where the scalar field is a monotonic function, varying 
from a maximal value at the centre to a minimum at spatial infinity. 
Thus there are no strong oscillations of the scalar field that would 
make the overall variation of the scalar field a bad indicator of the 
gradient terms.

\subsection{Spherically symmetric boson stars}
Let us start by considering spherically symmetric boson stars. These were first discussed by Kaup~\cite{Kaup:1968zz} and Ruffini and Bonazzola~\cite{Ruffini:1969qy} and were generalised by Colpi \textit{et al.}~\cite{Colpi:1986ye} for the model with self-interactions~(\ref{newL}). A fundamental property of boson stars, that allows circumventing Derrick-type arguments~\cite{Derrick:1964ww} (or virial identities) for the absence of stable, stationary solitonic solutions in a field theory, is the existence of a harmonic time dependence for the scalar field, 
\begin{equation}
\psi=f(r)e^{-iw t} \ ,
\end{equation}
for a spherical boson star. This time dependence vanishes at the level of the energy momentum tensor, due to the complex nature of the scalar field, making the scalar field ansatz compatible with a static line element. This introduces a frequency, $w$, as a fundamental parameter of each boson star solution. Above, $f(r)$ is a radial function obtained by numerically solving the field equations.

\begin{figure}[h!]
\begin{center}
\includegraphics[width=0.495\textwidth]{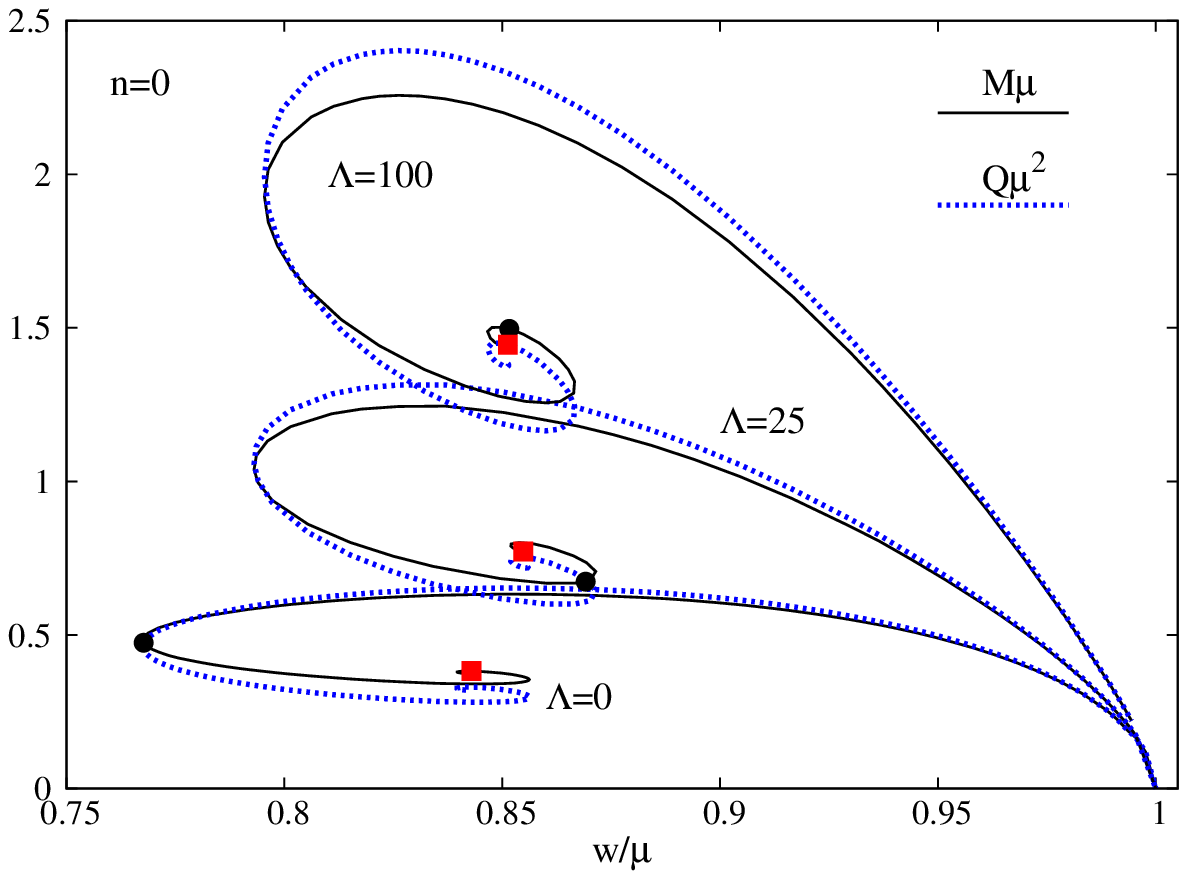}
\includegraphics[width=0.495\textwidth]{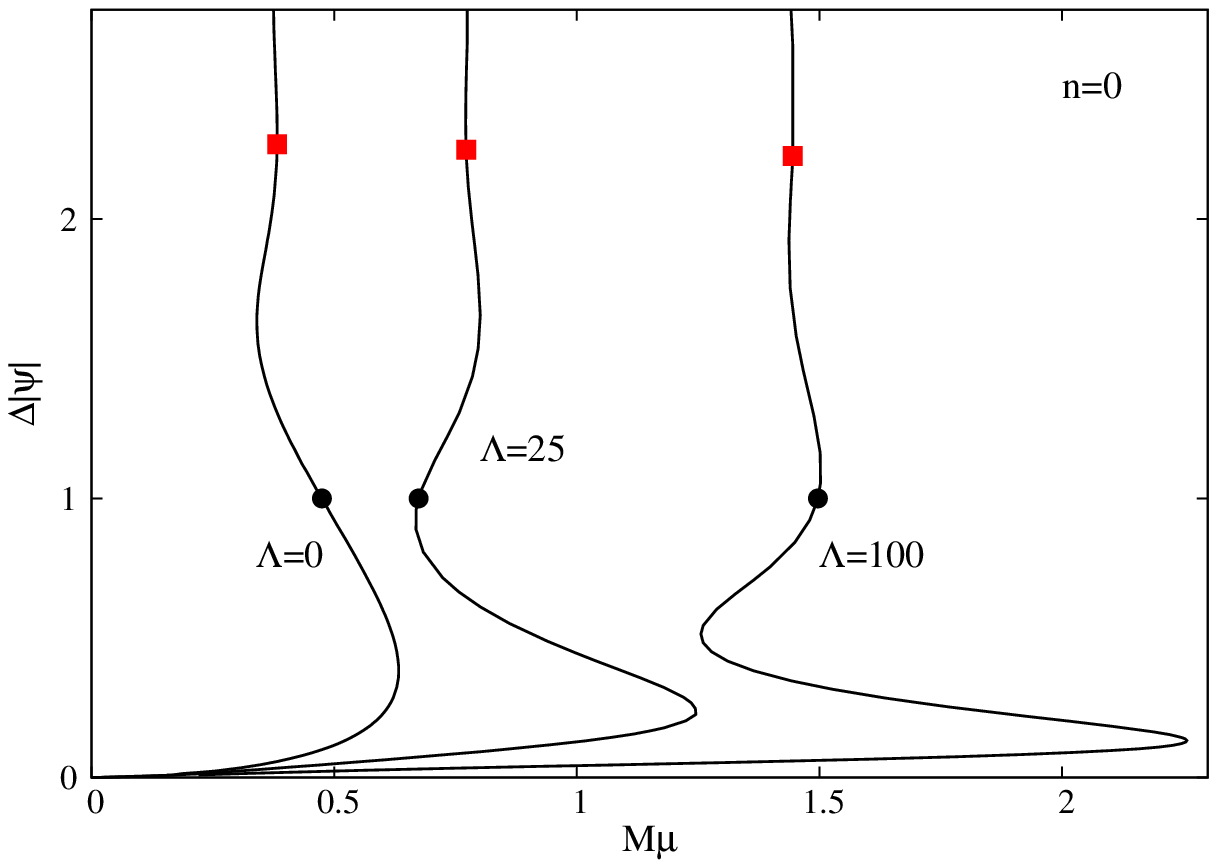} 
\caption{
\small{
{\it (Left panel)}
ADM mass $M$ or Noether charge $Q$ $vs.$ scalar field frequency $w$ diagram for static spherically symmetric boson stars with a quartic self-interaction and different values of the (rescaled) self-interaction coupling $\Lambda$, where $\Lambda= {\lambda}/{ \mu^2}$. 
{\it (Right panel)} Maximal variation of the scalar field as a function of the ADM mass for the same configurations.
In both plots, the black dots indicate where the swampland criteria start to fail,
while the red squares show where the
boson stars become ultra-compact.
 Only fundamental states are  shown (zero nodes, $n=0$).
}}
\label{fig1}
\end{center}
\end{figure}

 Spherical boson stars (as rotating ones) have fundamental states, for which the scalar field has no nodes ($n=0$, where $n$ is the number of nodes) and excited states with $n\neq 0$. The latter are generically unstable~\cite{Balakrishna:1997ej}, whereas the former have a branch of stable solutions. This stable branch, in an ADM mass, $M$, $vs.$ the scalar field frequency $w$ diagram, connects the solution with the maximal frequency with the solution with the maximal ADM mass - Figure~\ref{fig1}. Furthermore, spherical boson stars typically attain the maximum value of the scalar field at their centre $r=0$. Thus, since the scalar field vanishes at spatial infinity, for the swampland criterion, $\Delta |\psi|=|\psi|(0)=f(0)$. This value of the field is given, of course, in Planck units.

%
In Figure \ref{fig1} (left panel) we show the $M$ $vs.$ $w$ relation  for fundamental spherical boson stars, with various values of the self-interacting coupling (black solid lines). 
The typical pattern is a spiral, regardless of the value of $\Lambda$. As an intuition, the compactness of the boson stars roughly increases along the spiral, starting from the maximal frequency -- see~\cite{Cunha:2017wao,Herdeiro:2015gia,Herdeiro:2016gxs} for a precise plot. The black dots in the figure show the point at which the swampland criteria start to fail, $i.e$ the \textit{landscape-swampland transition}. Solutions from the maximal frequency $w=\mu$ up to this point, which always includes the stable branch, obey the criteria and thus can be part of the landscape. The red squares in the figure, on the other hand, show the point at which boson stars become ultra-compact; $i.e.$, they become sufficiently compact to develop a light right -- see also~\cite{Cunha:2017wao,Cunha:2015yba}. Light rings, therefore, always arise in the swampland region, rather than the landscape one. This figure also shows the Noether charge, $Q$ (blue dotted line), the conserved quantity associated to the global $U(1)$ symmetry of the complex scalar field. It can be interpreted as the number of scalar particles and, in units of $\mu$, its comparison with the ADM mass gives a sufficient (but not necessary) criterion for perturbative instability: solutions with $M>Q$ should be unstable against fission, as they have excess energy rather than binding energy. It can be seen from the figure that the landscape-swampland transition occurs also in the region where this fission instability is present, except for the $\Lambda=0$ case where both transitions roughly coincide. Note, however, that generic perturbative stability is more stringent than stability against fission. Figure \ref{fig1} (right panel) shows the same data but in a diagram where the value of the scalar field at the origin is plotted against the ADM mass.

\begin{figure}[h!]
\begin{center}
\includegraphics[width=0.495\textwidth]{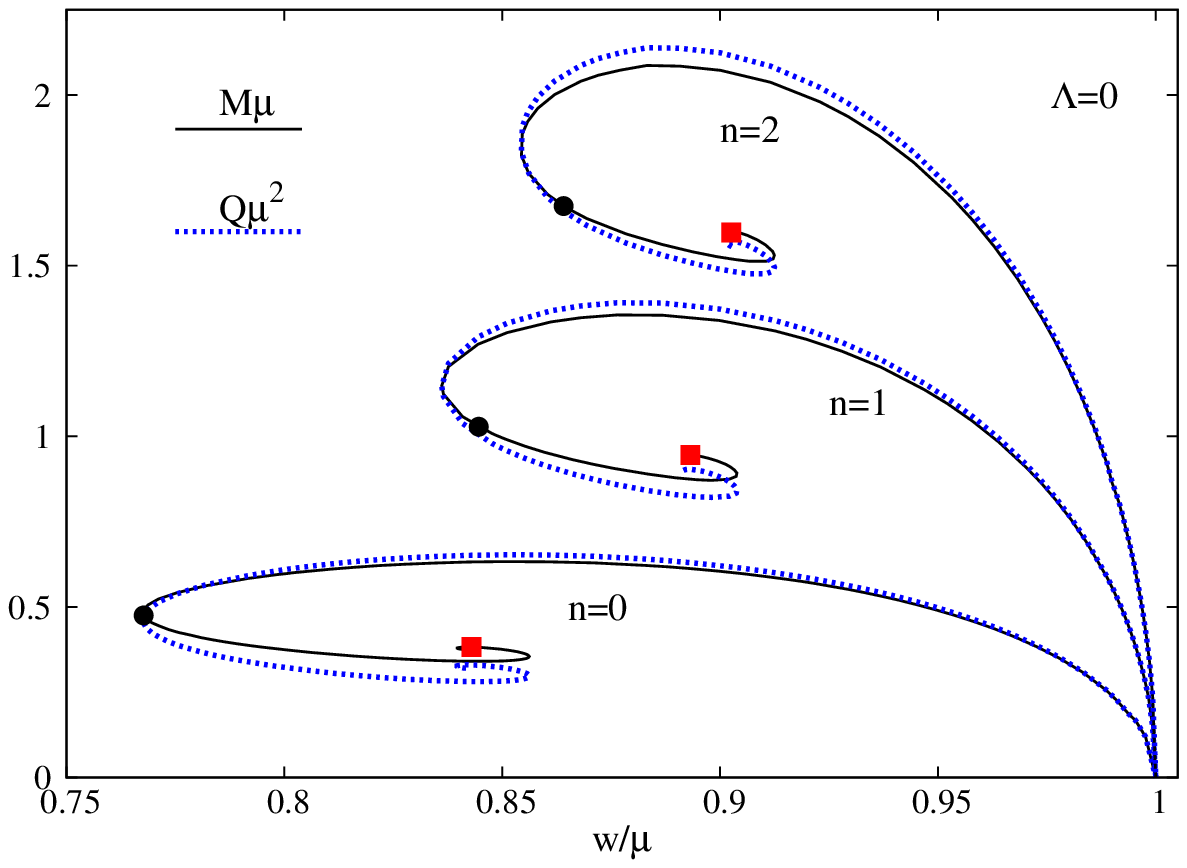}
\includegraphics[width=0.495\textwidth]{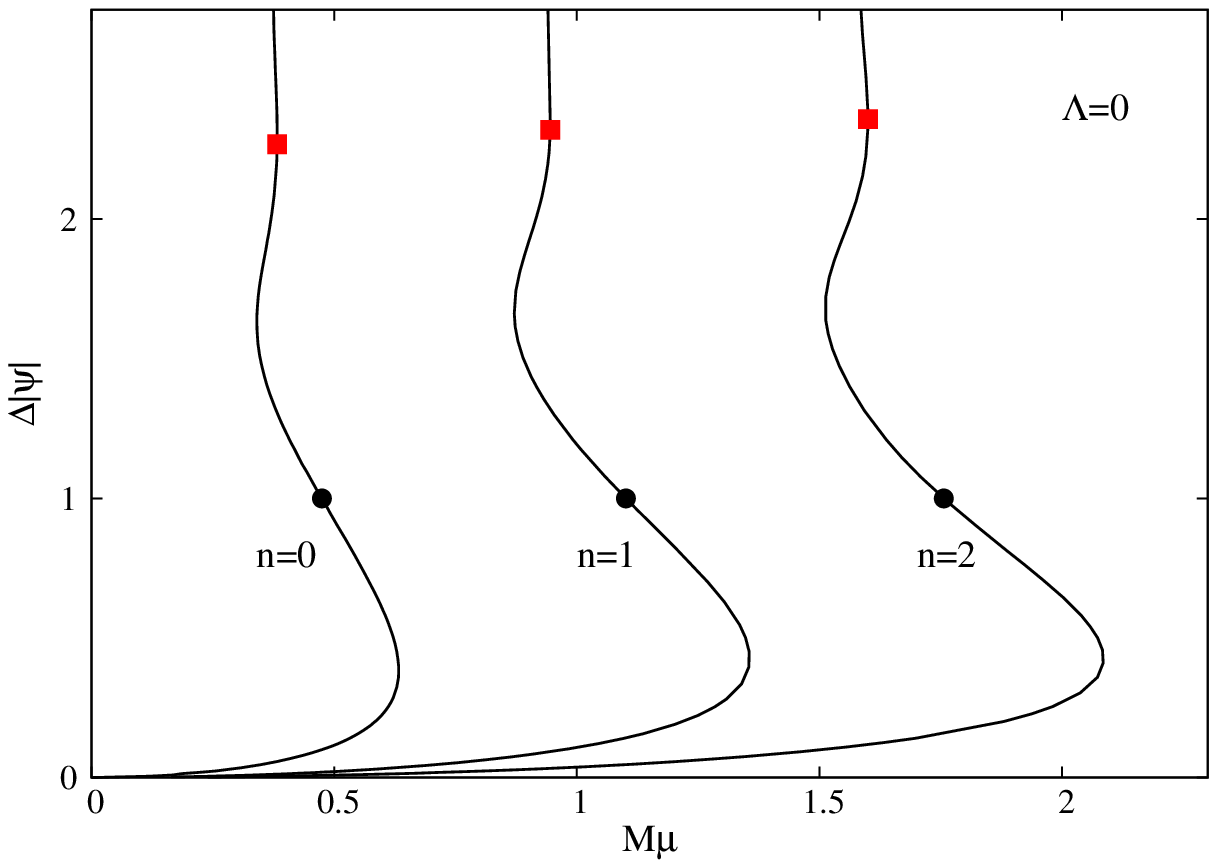} 
\caption{
\small{
Same as in Figure~1 but for excited states ($n\neq 0$) without self-interactions ($\Lambda=0$).}}
\label{fig2}
\end{center}
\end{figure}  
%
In Figure \ref{fig2} the same type of analysis done in Figure~\ref{fig1} is performed for excited states without self-interactions. Even though these excited states are not expected to have a stable branch, the pattern is similar. The landscape-swampland transition occurs beyond the maximum mass along the spiral, and  before light rings appear.

\subsection{Rotating boson stars and hairy black holes}
\begin{figure}[h!]
\begin{center}
\includegraphics[width=0.495\textwidth]{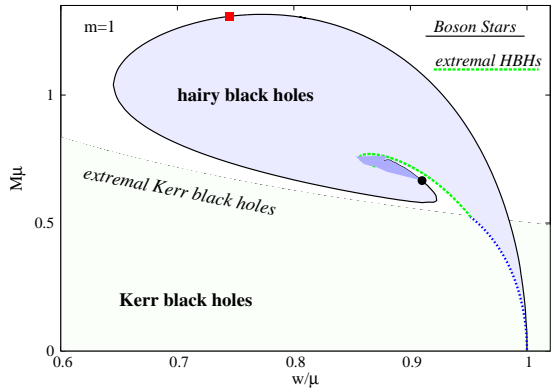} 
\caption{
\small{ 
Fundamental rotating boson star solutions with $m =  1$ (black solid line) in a mass $vs.$ scalar field frequency  
diagram (in units of $\mu$). The hairy black holes exist within
the blue shaded region. The swampland criteria fail to be satisfied inside the darker shaded region.
      }
}
\label{fig3}
\end{center}
\end{figure}

Including rotation, we observe both similar and different trends. In Figure  \ref{fig3} we show the ADM mass $vs.$ frequency relation for rotating boson stars of the same model with $\Lambda=0$ (red solid curve). For rotating boson stars, the scalar field has the form $\psi=g(r,\theta)e^{-i(wt-m\varphi)}$, in a spheroidal coordinate system $(t,r,\theta,\varphi)$, where $g(r,\theta)$ is a function determined by numerically solving the field equations - see $e.g.$~\cite{Herdeiro:2015gia} for details.  In the plot we exhibit solutions with the azimuthal harmonic index $m=1$; in general there are solution for each $m\in \mathbb{Z}$. Again, a stable branch against perturbations is expected to exist between the maximal frequency and the maximal ADM mass - see~\cite{Herdeiro:2015gia} for a discussion and references. The landscape-swampland transition (black dot as before)  always occurs well within the unstable branch. Here, however, the appearance of the light ring (red square as before) occurs before this transition, along the spiral. So there are ultra-compact rotating boson stars that are still in the landscape region. This shows ultra-compactness may still be compatible with the swampland criteria. Note, however, that for this model, such ultra-compact boson stars are always perturbatively unstable.

In Figure~\ref{fig3} we also show the domain of existence of black holes with synchronised hair~\cite{Herdeiro:2014goa} (blue shaded region), interpolating between this family of rotating boson stars (thick black solid line) and Kerr black holes (that exist below the thin solid line). Such hairy black holes can be seen as a bound state of a rotating horizon with a rotating boson star, a critical condition being the synchronous rotation between them, which allows circumventing various no scalar hair theorems for black holes - see~\cite{Herdeiro:2015waa} for a review. These hairy black holes reduce to some specific Kerr solutions in the limit of vanishing scalar field (blue dotted line), corresponding to Kerr black holes that can support stationary scalar clouds at the threshold of the 
superradiant instability~
\cite{Hod:2012px}. 
The domain of existence of hairy black holes has also a limit where they become extremal, $i.e.$ zero Hawing temperature (green dashed line). 

Our analysis shows that the domain where the swampland criteria fail are in the ``strong gravity region" at the centre of the domain of existence (dark shaded region) and continuously connected to the landscape-swampland transition point for the rotating boson stars.

%
\section{Discussion and remarks} 
\label{sec:discussion}

In this paper we have considered the swampland criteria recently proposed in a novel direction: to assess if vacua describing solitonic compact objects or hairy black holes may be part of the swampland or the string landscape.  
For simplicity, we have restricted to a model with two real scalar fields possessing the same mass and 
a positive quartic self-interacting term in the potential.

Perhaps the most interesting physical observation of the analysis presented herein is that in all boson star models studied so far, when the swampland criteria fail the solutions are unstable. Thus all stable solutions we have analysed are compatible with the landscape. It would be interesting to see how general this statement is by analysing more generic boson star models. We remark, however, that this and the other results in this discussion should be taken with the appropriate care, as the swampland criteria are only providing an estimate, whereas here we are taking the corresponding inequalities as strict. 

We have also considered ultra-compact solitonic objects, $i.e.$ boson stars with light rings. In this case the statement is model dependent. All static boson stars analysed can only be ultra-compact in the swampland. Rotating boson stars, on the other hand, can become ultra-compact and be part of the landscape. It would be quite interesting to find if there are models where such ultra-compact boson stars can be simultaneously part of the landscape and perturbatively stable, which is not the case for the rotating boson stars analysed here (with just a mass term in the potential). 

The conclusions in the last paragraph, namely that ultracompact spherical boson stars are in the swampland, depends crucially on taking the swampland conjecture as a 
strict inequality. A more relaxed criterion of the sort  $c\sim {\cal O}(\varepsilon^{-1})$\,, 
where $\varepsilon\ll 1$\,, would strengthen our main result, i.e. that stable configurations are in the landscape. In this sense our conclusions are conservative.  For such more relaxed criterion, we may even have spherical ultra-compact solutions in the landscape. 

We have also analysed some black hole solutions in equilibrium with a scalar field condensate around them - black holes with synchronised scalar hair. Here the region where the swampland conditions fail is the strong gravity region, which does not intersect with the region in the domain of existence where such black holes may be astrophysically viable~\cite{Degollado:2018ypf}.

Finally, let us remark that a (rather unexpected) 
feature unveiled by our study is that the two  swampland  criteria are not independent. 
That is, the criterion resulting from the requirement that the scalar field does not perform large excursions 
in field space suffices to obey both criteria. 
To some extent, this is a consequence of the specific choice
of the potential in (\ref{newL}). 
However, 
we have verified that this feature still holds 
even if higher order (positive) interaction terms  (of type $|\psi|^{2n}$, $n\geqslant 3$) are included in the model,
which also leads to the same pattern of the solutions.
 This interplay between the two criteria may change, however, 
for more complicated scalar field models, in particular those allowing for Q-balls 
(solitonic compact objects which exist even in the absence of gravity) 
or for axion stars (in a model with a gravitating single real scalar field) ~\cite{Schunck:2003kk}.  
Although these configurations share some basic features with boson stars, 
no general results can be established and  
a {\it case by case} analysis is required to clarify their status 
$w.r.t.$ the swampland criteria.

\section*{Acknowledgements}
This work has been supported by FCT (Portugal) through: the IF programme, grant PTDC/FIS-OUT/28407/2017, 
the  CIDMA strategic project UID/MAT/04106/2013 and the CENTRA strategic project UID/FIS/00099/2013. 
We also acknowledge support from  the  European  Union's  Horizon  2020  
research  and  innovation  (RISE) programmes H2020-MSCA-RISE-2015
Grant No.~StronGrHEP-690904 and H2020-MSCA-RISE-2017 Grant No.~FunFiCO-777740. 
The authors would like to acknowledge
networking support by the
COST Action CA16104. 
E.R. gratefully acknowledges the support of DIAS. K.U. is also supported by Grants-in-Aid from the
Scientific Research Fund of the Japan Society for the Promotion of
Science, under Contracts No. 16K05364.




\begin{thebibliography}{99}

\bibitem{Riess:1998cb}
  A.~G.~Riess {\it et al.} [Supernova Search Team],
  Astron.\ J.\  {\bf 116} (1998) 1009
  [astro-ph/9805201];
%
  S.~Perlmutter, M.~S.~Turner and M.~J.~White,
  Phys.\ Rev.\ Lett.\  {\bf 83} (1999) 670
  [astro-ph/9901052];
%
  C.~L.~Bennett {\it et al.} [WMAP Collaboration],
  Astrophys.\ J.\ Suppl.\  {\bf 208} (2013) 20
  [arXiv:1212.5225 [astro-ph.CO]];
%
  P.~A.~R.~Ade {\it et al.} [Planck Collaboration],
  Astron.\ Astrophys.\  {\bf 571} (2014) A16
  [arXiv:1303.5076 [astro-ph.CO]];
%
  C.~Blake and K.~Glazebrook,
  Astrophys.\ J.\  {\bf 594} (2003) 665
  [astro-ph/0301632];
%
  H.~J.~Seo and D.~J.~Eisenstein,
  Astrophys.\ J.\  {\bf 598} (2003) 720
  [astro-ph/0307460]%
%
  S.~D.~M.~White, J.~F.~Navarro, A.~E.~Evrard and C.~S.~Frenk,
  Nature {\bf 366} (1993) 429%
%
  P.~Schuecker, H.~Bohringer, C.~A.~Collins and L.~Guzzo,
  Astron.\ Astrophys.\  {\bf 398} (2003) 867
  [astro-ph/0208251];
%
  M.~Kilbinger {\it et al.},
  Astron.\ Astrophys.\  {\bf 497} (2009) 677
  [arXiv:0810.5129 [astro-ph]];
%
  D.~M.~Scolnic {\it et al.},
  Astrophys.\ J.\  {\bf 859} (2018) no.2,  101
  [arXiv:1710.00845 [astro-ph.CO]]%
%
  N.~Aghanim {\it et al.} [Planck Collaboration],
  arXiv:1807.06209 [astro-ph.CO];
%
  P.~A.~R.~Ade {\it et al.} [BICEP2 and Keck Array Collaborations],
  Astrophys.\ J.\  {\bf 811} (2015) 126
  [arXiv:1502.00643 [astro-ph.CO]].
\bibitem{Grana:2005jc}
  M.~Grana,
  Phys.\ Rept.\  {\bf 423} (2006) 91
  [hep-th/0509003].

\bibitem{Douglas:2006es}
  M.~R.~Douglas and S.~Kachru,
  Rev.\ Mod.\ Phys.\  {\bf 79} (2007) 733
  [hep-th/0610102].

\bibitem{Blumenhagen:2006ci}
  R.~Blumenhagen, B.~Kors, D.~Lust and S.~Stieberger,
  Phys.\ Rept.\  {\bf 445} (2007) 1
  [hep-th/0610327].

\bibitem{Gukov:1999ya}
  S.~Gukov, C.~Vafa and E.~Witten,
  Nucl.\ Phys.\ B {\bf 584} (2000) 69
   Erratum: [Nucl.\ Phys.\ B {\bf 608} (2001) 477]
  [hep-th/9906070];
%
  S.~B.~Giddings, S.~Kachru and J.~Polchinski,
  Phys.\ Rev.\ D {\bf 66} (2002) 106006
  [hep-th/0105097];
%
  S.~Kachru, M.~B.~Schulz and S.~Trivedi,
  JHEP {\bf 0310} (2003) 007
  [hep-th/0201028].

\bibitem{Kachru:2003aw}
  S.~Kachru, R.~Kallosh, A.~D.~Linde and S.~P.~Trivedi,
  Phys.\ Rev.\ D {\bf 68} (2003) 046005
  [hep-th/0301240].

\bibitem{Kachru:2003sx}
  S.~Kachru, R.~Kallosh, A.~D.~Linde, J.~M.~Maldacena, L.~P.~McAllister and S.~P.~Trivedi,
  JCAP {\bf 0310} (2003) 013
  [hep-th/0308055].

\bibitem{Balasubramanian:2005zx}
  V.~Balasubramanian, P.~Berglund, J.~P.~Conlon and F.~Quevedo,
  JHEP {\bf 0503} (2005) 007
  [hep-th/0502058];
%
  H.~Kodama and K.~Uzawa,
  JHEP {\bf 0507} (2005) 061
  [hep-th/0504193];
%
  O.~DeWolfe, A.~Giryavets, S.~Kachru and W.~Taylor,
  JHEP {\bf 0507} (2005) 066
  [hep-th/0505160];
%
  H.~Kodama and K.~Uzawa,
  JHEP {\bf 0603} (2006) 053
  [hep-th/0512104].
\bibitem{deWit:1986mwo}
  B.~de Wit, D.~J.~Smit and N.~D.~Hari Dass,
  Nucl.\ Phys.\ B {\bf 283} (1987) 165;
%
\bibitem{Maldacena:2000mw}
  J.~M.~Maldacena and C.~Nunez,
  Int.\ J.\ Mod.\ Phys.\ A {\bf 16} (2001) 822
  [hep-th/0007018];
%
\bibitem{Ivanov:2000fg}
  S.~Ivanov and G.~Papadopoulos,
  Phys.\ Lett.\ B {\bf 497} (2001) 309
  [hep-th/0008232].
\bibitem{Goheer:2002vf}
  N.~Goheer, M.~Kleban and L.~Susskind,
  JHEP {\bf 0307} (2003) 056
  [hep-th/0212209];
%
  M.~P.~Hertzberg, S.~Kachru, W.~Taylor and M.~Tegmark,
  JHEP {\bf 0712} (2007) 095
  [arXiv:0711.2512 [hep-th]].
 
 \bibitem{Danielsson:2018ztv}
  U.~H.~Danielsson and T.~Van Riet,
  Int.\ J.\ Mod.\ Phys.\ D {\bf 27} (2018) no.12,  1830007
  [arXiv:1804.01120 [hep-th]].
\bibitem{Susskind:2003kw}
  L.~Susskind,
  in Carr, Bernard (ed.): {\it  Universe or multiverse?} 247-266
  [hep-th/0302219];
%
  T.~Banks, M.~Dine and E.~Gorbatov,
  JHEP {\bf 0408} (2004) 058
  [hep-th/0309170];
%
  F.~Denef and M.~R.~Douglas,
  JHEP {\bf 0405} (2004) 072
  [hep-th/0404116];
%
  R.~Kallosh and A.~D.~Linde,
  JHEP {\bf 0412} (2004) 004
  [hep-th/0411011].
\bibitem{Vafa:2005ui}
  C.~Vafa,
  hep-th/0509212.

\bibitem{Denef:2008wq}
  F.~Denef,
  Les Houches {\bf 87} (2008) 483
  [arXiv:0803.1194 [hep-th]].

\bibitem{Brennan:2017rbf}
  T.~D.~Brennan, F.~Carta and C.~Vafa,
  PoS TASI {\bf 2017} (2017) 015
  [arXiv:1711.00864 [hep-th]].

\bibitem{Ooguri:2006in}
  H.~Ooguri and C.~Vafa,
  Nucl.\ Phys.\ B {\bf 766} (2007) 21
  [hep-th/0605264].
\bibitem{Ooguri:2016pdq}
  H.~Ooguri and C.~Vafa,
  Adv.\ Theor.\ Math.\ Phys.\  {\bf 21} (2017) 1787
  [arXiv:1610.01533 [hep-th]].
   
\bibitem{Obied:2018sgi}
  G.~Obied, H.~Ooguri, L.~Spodyneiko and C.~Vafa,
  arXiv:1806.08362 [hep-th].
 
 \bibitem{Agrawal:2018own}
  P.~Agrawal, G.~Obied, P.~J.~Steinhardt and C.~Vafa,
  Phys.\ Lett.\ B {\bf 784} (2018) 271
  [arXiv:1806.09718 [hep-th]].
 
 \bibitem{Ooguri:2018wrx}
  H.~Ooguri, E.~Palti, G.~Shiu and C.~Vafa,
  Phys.\ Lett.\ B {\bf 788} (2019) 180
  [arXiv:1810.05506 [hep-th]].
\bibitem{Brown:2015iha}
  J.~Brown, W.~Cottrell, G.~Shiu and P.~Soler,
  JHEP {\bf 1510} (2015) 023
  [arXiv:1503.04783 [hep-th]];
  %
  R.~Blumenhagen, I.~Valenzuela and F.~Wolf,
  JHEP {\bf 1707} (2017) 145
  [arXiv:1703.05776 [hep-th]];
%
  A.~Ach\'ucarro and G.~A.~Palma,
  arXiv:1807.04390 [hep-th];
%
  S.~K.~Garg and C.~Krishnan,
  arXiv:1807.05193 [hep-th];
%
  A.~Kehagias and A.~Riotto,
  Fortsch.\ Phys.\  {\bf 66} (2018) no.10,  1800052
  [arXiv:1807.05445 [hep-th]];
%
  H.~Matsui and F.~Takahashi,
  Phys.\ Rev.\ D {\bf 99} (2019) 023533 
  [arXiv:1807.11938 [hep-th]];
  %
  C.~Damian and O.~Loaiza-Brito,
  arXiv:1808.03397 [hep-th];
%
  W.~H.~Kinney, S.~Vagnozzi and L.~Visinelli,
  arXiv:1808.06424 [astro-ph.CO];
%
  S.~Brahma and M.~Wali Hossain,
  arXiv:1809.01277 [hep-th];
%
  C.~Han, S.~Pi and M.~Sasaki,
  arXiv:1809.05507 [hep-ph];
%
  K.~Dimopoulos,
   Phys.\ Rev.\ D {\bf 98} (2018) no.12,  123516
  [arXiv:1810.03438 [gr-qc]].;
%
  C.~M.~Lin, K.~W.~Ng and K.~Cheung,
  arXiv:1810.01644 [hep-ph];
%
  A.~Ashoorioon,
  arXiv:1810.04001 [hep-th].
%
  S.~Das,
  arXiv:1810.05038 [hep-th];
%
  S.~J.~Wang,
  Phys.\ Rev.\ D {\bf 99} (2019) 023529 
  [arXiv:1810.06445 [hep-th]];
%
  S.~K.~Garg, C.~Krishnan and M.~Zaid,
  arXiv:1810.09406 [hep-th];
%
  J.~J.~Heckman, C.~Lawrie, L.~Lin and G.~Zoccarato,
  arXiv:1811.01959 [hep-th];
%
  C.~I.~Chiang, J.~M.~Leedom and H.~Murayama,
  arXiv:1811.01987 [hep-th].
\bibitem{Schunck:2003kk}
  F.~E.~Schunck and E.~W.~Mielke,
  Class.\ Quant.\ Grav.\  {\bf 20} (2003) R301
  [arXiv:0801.0307 [astro-ph]];
  %
  S.~L.~Liebling and C.~Palenzuela,
  Living Rev.\ Rel.\  {\bf 15} (2012) 6
   [Living Rev.\ Rel.\  {\bf 20} (2017) no.1,  5]
  [arXiv:1202.5809 [gr-qc]].
\bibitem{Kaup:1968zz}
  D.~J.~Kaup,
  Phys.\ Rev.\  {\bf 172} (1968) 1331.
  
\bibitem{Ruffini:1969qy}
  R.~Ruffini and S.~Bonazzola,
  Phys.\ Rev.\  {\bf 187} (1969) 1767.
\bibitem{Seidel:1993zk}
  E.~Seidel and W.~M.~Suen,
  Phys.\ Rev.\ Lett.\  {\bf 72} (1994) 2516
  [gr-qc/9309015].
\bibitem{Jetzer:1988vr}
  P.~Jetzer,
  Nucl.\ Phys.\ B {\bf 316} (1989) 411;
%
  M.~Gleiser and R.~Watkins,
  Nucl.\ Phys.\ B {\bf 319} (1989) 733;
%
  T.~D.~Lee and Y.~Pang,
  Nucl.\ Phys.\ B {\bf 315} (1989) 477.

\bibitem{Guzman:2009zz}
  F.~S.~Guzman and J.~M.~Rueda-Becerril,
  Phys.\ Rev.\ D {\bf 80} (2009) 084023
  [arXiv:1009.1250 [astro-ph.HE]];
%
  F.~H.~Vincent, Z.~Meliani, P.~Grandclement, E.~Gourgoulhon and O.~Straub,
  Class.\ Quant.\ Grav.\  {\bf 33} (2016) no.10,  105015
  [arXiv:1510.04170 [gr-qc]].

\bibitem{Cunha:2017wao}
  P.~V.~P.~Cunha, J.~A.~Font, C.~Herdeiro, E.~Radu, N.~Sanchis-Gual and M.~Zilhao,
  Phys.\ Rev.\ D {\bf 96} (2017) no.10,  104040
  [arXiv:1709.06118 [gr-qc]].
%
\bibitem{Suarez:2013iw}
  A.~Sumgez, V.~H.~Robles and T.~Matos,
  Astrophys.\ Space Sci.\ Proc.\  {\bf 38} (2014) 107
  [arXiv:1302.0903 [astro-ph.CO]];
  %
\bibitem{Krippendorf:2018tei}
  S.~Krippendorf, F.~Muia and F.~Quevedo,
  JHEP {\bf 1808} (2018) 070
  [arXiv:1806.04690 [hep-th]].
%
  B.~Li, T.~Rindler-Daller and P.~R.~Shapiro,
  Phys.\ Rev.\ D {\bf 89} (2014) no.8,  083536
  [arXiv:1310.6061 [astro-ph.CO]].
\bibitem{Schunck:1996he}
  F.~E.~Schunck and E.~W.~Mielke,
  Phys.\ Lett.\ A {\bf 249} (1998) 389;
%
  S.~Yoshida and Y.~Eriguchi,
  Phys.\ Rev.\ D {\bf 56} (1997) 762.
\bibitem{Herdeiro:2014goa}
  C.~A.~R.~Herdeiro and E.~Radu,
  Phys.\ Rev.\ Lett.\  {\bf 112} (2014) 221101
  [arXiv:1403.2757 [gr-qc]].

\bibitem{Herdeiro:2015gia}
  C.~Herdeiro and E.~Radu,
  Class.\ Quant.\ Grav.\  {\bf 32} (2015) no.14,  144001
  [arXiv:1501.04319 [gr-qc]].
  
\bibitem{Kerr:1963ud}
  R.~P.~Kerr,
  Phys.\ Rev.\ Lett.\  {\bf 11} (1963) 237.
  
\bibitem{Colpi:1986ye}
  M.~Colpi, S.~L.~Shapiro and I.~Wasserman,
  Phys.\ Rev.\ Lett.\  {\bf 57} (1986) 2485.
   
\bibitem{Cardoso:2016rao}
  V.~Cardoso, E.~Franzin and P.~Pani,
  Phys.\ Rev.\ Lett.\  {\bf 116} (2016) no.17,  171101
   Erratum: [Phys.\ Rev.\ Lett.\  {\bf 117} (2016) no.8,  089902]
  [arXiv:1602.07309 [gr-qc]];
  %
  P.~V.~P.~Cunha, E.~Berti and C.~A.~R.~Herdeiro,
  Phys.\ Rev.\ Lett.\  {\bf 119} (2017) no.25,  251102
  [arXiv:1708.04211 [gr-qc]].

\bibitem{Dasgupta:2018rtp}
  K.~Dasgupta, M.~Emelin, E.~McDonough and R.~Tatar,
  JHEP {\bf 1901} (2019) 145 
  [arXiv:1808.07498 [hep-th]];
%
  U.~Danielsson,
  arXiv:1809.04512 [hep-th].
\bibitem{Grimm:2018ohb}
  T.~W.~Grimm, E.~Palti and I.~Valenzuela,
  JHEP {\bf 1808} (2018) 143
  [arXiv:1802.08264 [hep-th]];
%
  R.~Blumenhagen,
  PoS CORFU {\bf 2017} (2018) 175
  [arXiv:1804.10504 [hep-th]].

\bibitem{Andriot:2018mav}
  D.~Andriot and C.~Roupec,
  arXiv:1811.08889 [hep-th].

\bibitem{Denef:2018etk}
  F.~Denef, A.~Hebecker and T.~Wrase,
  Phys.\ Rev.\ D {\bf 98} (2018) no.8,  086004
  [arXiv:1807.06581 [hep-th]];
%
  J.~P.~Conlon,
  Int.\ J.\ Mod.\ Phys.\ A {\bf 33} (2018) no.29,  1850178
  [arXiv:1808.05040 [hep-th]];
%
  H.~Murayama, M.~Yamazaki and T.~T.~Yanagida,
  JHEP {\bf 1812} (2018) 032
  [arXiv:1809.00478 [hep-th]].;
%
  K.~Choi, D.~Chway and C.~S.~Shin,
  JHEP {\bf 1811} (2018) 142
  [arXiv:1809.01475 [hep-th]].;
%
  K.~Hamaguchi, M.~Ibe and T.~Moroi,
   JHEP {\bf 1812} (2018) 023
  [arXiv:1810.02095 [hep-th]].;
%
  M.~Emelin and R.~Tatar,
  arXiv:1811.07378 [hep-th].
%
\bibitem{Blaback:2018hdo}
  J.~Bl\aa b\"ack, U.~Danielsson and G.~Dibitetto,
  arXiv:1810.11365 [hep-th].


\bibitem{Arvanitaki:2009fg}
  A.~Arvanitaki, S.~Dimopoulos, S.~Dubovsky, N.~Kaloper and J.~March-Russell,
  Phys.\ Rev.\ D {\bf 81} (2010) 123530
  [arXiv:0905.4720 [hep-th]].

  
\bibitem{Herdeiro:2015tia}
  C.~A.~R.~Herdeiro, E.~Radu and H.~Runarsson,
  Phys.\ Rev.\ D {\bf 92} (2015) no.8,  084059
  [arXiv:1509.02923 [gr-qc]].


\bibitem{Derrick:1964ww}
  G.~H.~Derrick,
  J.\ Math.\ Phys.\  {\bf 5} (1964) 1252.

\bibitem{Balakrishna:1997ej}
  J.~Balakrishna, E.~Seidel and W.~M.~Suen,
  Phys.\ Rev.\ D {\bf 58} (1998) 104004
  [gr-qc/9712064].
  
\bibitem{Herdeiro:2016gxs}
  C.~A.~R.~Herdeiro, E.~Radu and H.~F.~Runarsson,
  Int.\ J.\ Mod.\ Phys.\ D {\bf 25} (2016) no.09,  1641014
  [arXiv:1604.06202 [gr-qc]].
  
\bibitem{Cunha:2015yba}
  P.~V.~P.~Cunha, C.~A.~R.~Herdeiro, E.~Radu and H.~F.~Runarsson,
  Phys.\ Rev.\ Lett.\  {\bf 115} (2015) no.21,  211102
  [arXiv:1509.00021 [gr-qc]].

\bibitem{Herdeiro:2015waa}
  C.~A.~R.~Herdeiro and E.~Radu,
  Int.\ J.\ Mod.\ Phys.\ D {\bf 24} (2015) no.09,  1542014
  [arXiv:1504.08209 [gr-qc]].
\bibitem{Hod:2012px}
  S.~Hod,
  Phys.\ Rev.\ D {\bf 86} (2012) 104026
   Erratum: [Phys.\ Rev.\ D {\bf 86} (2012) 129902]
  [arXiv:1211.3202 [gr-qc]];
  %
  S.~Hod,
  Eur.\ Phys.\ J.\ C {\bf 73} (2013) no.4,  2378
  [arXiv:1311.5298 [gr-qc]];
 %
  S.~Hod,
  Phys.\ Rev.\ D {\bf 90} (2014) no.2,  024051
  [arXiv:1406.1179 [gr-qc]];
%
  C.~L.~Benone, L.~C.~B.~Crispino, C.~Herdeiro and E.~Radu,
  Phys.\ Rev.\ D {\bf 90} (2014) no.10,  104024
  [arXiv:1409.1593 [gr-qc]];
%
  S.~Hod,
  JHEP {\bf 1701} (2017) 030
  [arXiv:1612.00014 [hep-th]].
\bibitem{Degollado:2018ypf}
  J.~C.~Degollado, C.~A.~R.~Herdeiro and E.~Radu,
  Phys.\ Lett.\ B {\bf 781} (2018) 651
  [arXiv:1802.07266 [gr-qc]].

	
	
	
\end{thebibliography}
\end{document}